\documentclass[12pt]{iopart}
\usepackage{graphicx}
\usepackage{amssymb}
\usepackage[T1]{fontenc}
\usepackage[utf8]{inputenc}
\usepackage[usenames,dvipsnames]{color} 
\usepackage{array}
\usepackage{soul}

\begin{document}

\title{Suspended graphene devices with local gate control on an insulating substrate}

\author{Florian R Ong$^1$\footnote{Present address: Universit\"{a}t Innsbruck, Institut f\"{u}r Experimentalphysik, Technikerstrasse 25/4, A-6020 Innsbruck Austria},
Zheng Cui$^1$,
Muhammet A Yurtalan$^2$,
Cameron Vojvodin$^1$,
Michał Papaj  $^1$,
Jean-Luc F X Orgiazzi$^2$,
Chunqing Deng$^1$,
Mustafa Bal$^1$\footnote{Present address: TUBITAK Marmara Research Centre, Materials Institute, P.O. Box 21, 41470 Gebze, Kocaeli, Turkey}
and Adrian Lupascu$^1$}

\vspace{0.2cm}

\address{$^1$ Institute for Quantum Computing, Department of Physics and Astronomy, and Waterloo Institute for Nanotechnology,
University of Waterloo, Waterloo, Ontario N2L 3G1, Canada}

\vspace{0.2cm}

\address{$^2$ Institute for Quantum Computing, Department of Electrical Engineering, and Waterloo Institute for Nanotechnology, University of Waterloo, Waterloo, Ontario N2L 3G1, Canada}

\vspace{0.2cm}

\eads{\mailto{florian.ong@uibk.ac.at}, \mailto{alupascu@uwaterloo.ca}}

\begin{abstract}
We present a fabrication process for graphene-based devices where a graphene monolayer is suspended above a local metallic gate placed in a trench. As an example we detail the fabrication steps of a graphene field-effect transistor. The devices are built on a bare high-resistivity silicon substrate. At temperatures of 77~K and below, we observe the field-effect modulation of the graphene resistivity by a voltage applied to the gate. This fabrication approach enables new experiments involving graphene-based superconducting qubits and nano-electromechanical resonators. The method is applicable to other two-dimensional materials.
\end{abstract}


\vspace{2pc}

\submitto{\NT}

\maketitle

\section{Introduction}

Graphene physics has become a broad and very active field of research since single-layer graphene sheets have been successfully isolated ten years ago \cite{novoselov2004}. On the one hand, exceptional structural properties, such as two-dimensional crystalline order over macroscopic scale, make graphene a rich playground for investigating fundamental physics problems. Graphene research topics include dynamics of massless Dirac fermions \cite{novoselov2005}, quantum Hall effect \cite{zhang2005}, Klein paradox \cite{stander2009}, conductance quantization \cite{peres2006}, quantum billiards \cite{miao2007}, and superconducting proximity effect \cite{heersche2007}. On the other hand, the potential applications of graphene in micro- and nanotechnologies have triggered intense interest \cite{berger2004ultrathin,geim2007}. For instance, the outstanding electrical and mechanical properties of graphene could be the basis for next-generation transistors \cite{schwierz2010,weiss2012} and for classical and quantum nano-electromechanical (NEM) devices \cite{poot2012,chen2013}. Graphene is a versatile material allowing 2D electronic transport to be investigated in different regimes. By patterning devices with a length of the conduction channel varying by orders of magnitude (a few tens of nanometers up to centimeters), one easily explores the ballistic and diffusive regimes of electronic transport. Moreover, applying an electric field in the vicinity of a graphene sheet allows to change the nature of the charge carriers (electrons or holes) and to modulate their surface density. This offers a new degree of freedom compared to metals or 2D electron gases in GaAs/GaAlAs heterostructures.

A convenient method to study electron transport in graphene is based on the field effect transistor (FET) geometry, where a graphene sheet is galvanically connected to two metallic electrodes and capacitively coupled to a metallic gate \cite{novoselov2004}. The easiest and most common route of fabricating such devices relies on a doped semiconducting substrate covered by an insulating layer. The substrate, typically doped silicon, is used as a back-gate to modulate the density of carriers in graphene. Although this method and its refinements have proven very successful, in practice it has the following limitations in investigating the properties of graphene. Firstly, there is evidence that the intrinsic properties of graphene are impaired when directly lying on a bulk material (the insulating layer in this case). Interactions with the substrate introduce a scattering mechanism for electrons travelling through the conducting channel which reduces their mobility \cite{sabio2008,du2008} (a noteworthy exception is boron nitride substrate, onto which graphene performs well due to the matching of both hexagonal lattices \cite{dean2010}). Secondly, trapped charges in the oxide locally modify the electrical potential set by the back-gate \cite{tan2007}. Lastly, the use of the substrate as a back-gate makes it difficult to integrate several graphene devices on the same chip, for which multiple local gates are required.

The presence of a doped substrate is also detrimental to systems sensitive to energy losses. This includes superconducting qubits and electromagnetic resonators operating in the GHz frequency range. Superconductor-graphene-superconductor (S-G-S) junctions have been shown to exhibit Josephson tunneling with a critical current tunable by the gate voltage \cite{heersche2007,jeong2011,titov2006}. In a S-G-S junction the tunnel barrier is formed by a graphene sheet contacted laterally by superconducting electrodes, in a configuration similar to that of long supeconductor-normal metal-supeconductor junctions \cite{dubos2001josephson}. S-G-S junctions could be used as building blocks for superconducting qubits, SQUIDs \cite{lee2011,girit2008}, and superconducting circuits to study quantum optics at microwave frequencies \cite{singh2014}. In state-of-the-art superconducting qubits, the amorphous nature of the oxide forming the Josephson junctions sets a limit on performances: structural defects in the barriers result in fluctuating electric and magnetic fields that couple to the qubit and impair its coherence \cite{martinis2005}. Replacing the amorphous oxide with a well-ordered material like graphene has the potential to improve the coherence times of superconducting qubits. However, qubits with long coherence times as well as microwave resonators with high quality factors are produced on insulating substrates such as high-resistivity silicon or sapphire \cite{megrant2012,orgiazzi2014}. In this context, using a doped silicon substrate to gate a graphene Josephson junction must be avoided since its high density of mobile carriers leads to unacceptable radiofrequency losses. This problem may also be of concern in graphene NEMs operating in the GHz frequency range.

\begin{figure}[h]
\begin{center}
\includegraphics[width=14cm]{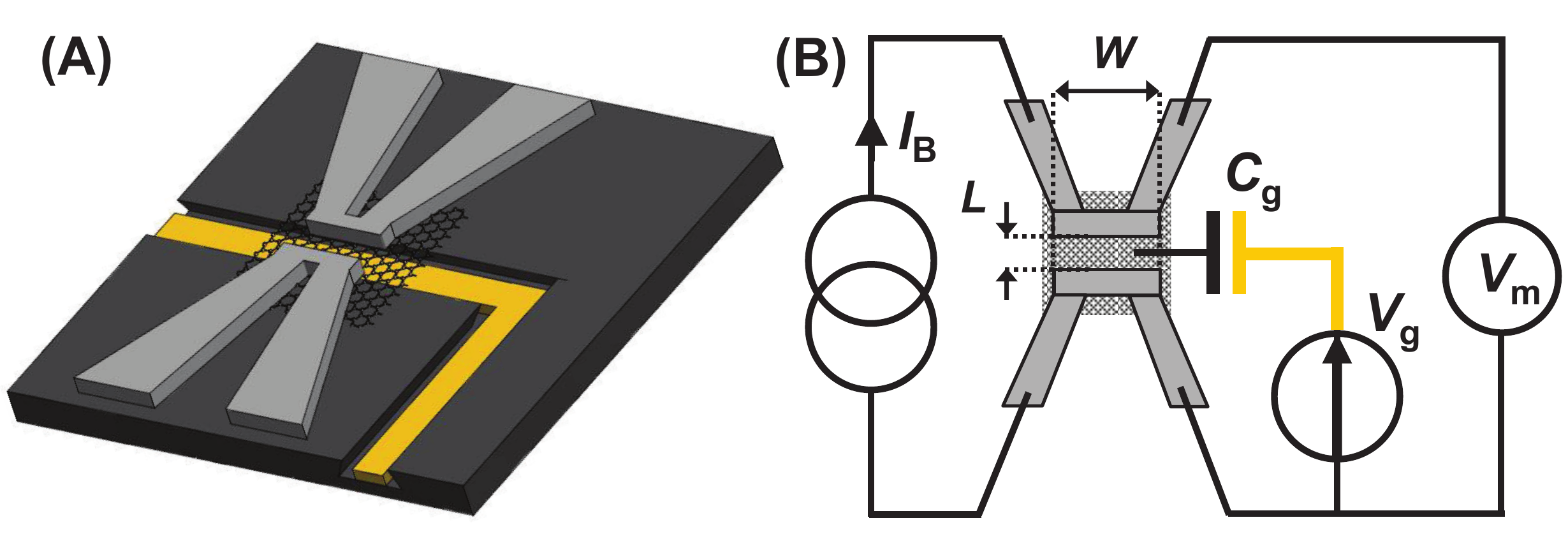}
\vspace{0cm}
\caption{ (color online) Overview of a suspended graphene junction. (A) Cartoon picture of the device (not to scale): a trench is etched in the silicon substrate (dark gray). A local gate (yellow) is deposited at the bottom of the trench. The graphene monolayer (honeycomb pattern) is suspended above the gate. Electrodes (gray) are evaporated to contact the graphene. (B) Sketch of the electrical setup to characterize the field-effect properties of the device, as discussed in the main text.}
\label{fig-sketches}
\end{center}
\end{figure}

In this article we present a fabrication method that circumvents the aforementioned drawbacks of graphene devices fabricated on doped silicon substrate. Some of the problems have already been addressed separately. Local gates have been implemented \cite{schwierz2010,liao2010b,grushina2013}. FETs based on suspended graphene have been successfully fabricated, preventing interaction with the substrate and demonstrating an enhanced mobility \cite{lau2012}. However, to the best of our knowledge, there is no report yet of a suspended graphene FET with a local gate built on an insulating substrate. We note that two recent publications \cite{castellanos2014, weber2014} describe a fabrication process that shares common features with our own. However, our approach to fabrication of suspended graphene with local gating includes a robust and well established recipe for low contact resistances. While designs in Refs. \cite{castellanos2014, weber2014} appear to be robust against contact resistance, other resonator designs may require lower contact resistances. Low contact resistances are essential in view of realizing  S-G-S Josephson junctions \cite{ojeda2009}.

A schematic picture of our suspended graphene junction is shown in \fref{fig-sketches}(A). The substrate is unoxidized high-resistivity silicon. A trench is etched in the silicon. A metallic electrode is deposited at the bottom of this trench to be used as a gate.  A graphene monolayer is suspended over the trench. The gate is separated from the graphene only by air (or vacuum). The ohmic contacts are deposited over the graphene sheet at the edge of the trench, ensuring that the whole surface of the conduction channel --- called junction all along this paper --- is suspended. The conductance of the device (including that of the junction and its ohmic contacts) is measured as a function of gate voltage $V_{\rm g}$ in a 4-point configuration, as shown in \fref{fig-sketches}(B). The devices investigated in this work have a short ($L$~=~350~-~500~nm) and wide ($W$~=~4~-~7~$\mu$m ) conduction channel, exhibiting typical graphene FET behaviour at low temperatures (4 to 77~K). The performance is comparable to that of unsuspended devices, with mobilities around 10,000~cm$^2$/V/s at a carrier density of $n=10^{11}$~cm$^{-2}$. However, the charge degeneracy point is found to be closer to zero gate voltage than that of typical graphene FETs lying on oxide, as expected and previously demonstrated for other types of suspended devices \cite{bolotin2008}.

The graphene junctions we introduce in this work satisfy all the requirements of high-transparency graphene Josephson junctions that could be used as building blocks for superconducting qubits. Indeed, compared to other recent implementations of suspended junctions \cite{mizuno2013}, our device lies on an insulating susbtrate, an important requirement for quantum information processing based on superconducting qubits. Moreover, the junctions could yield a well-controlled environment to investigate noise properties of transport in graphene \cite{zhang2011,balandin2013}. Finally, the fabrication process we introduce here could be adapted to fabricate nano-electromechanical resonators where a vibrating suspended graphene bridge or membrane is actuated at high frequency by a local gate \cite{bunch2007}.

The paper is structured as follows: in Section~2 we describe the fabrication process of the device in detail. In Section~3 we present field-effect measurements carried out on a typical device at temperatures ranging from 4~K to 300~K. At room temperature, gate leakages through the imperfectly insulating substrate degrade the FET behaviour. At low temperature these leakages are suppressed and the device performs normally. We provide a model for the gate leakages, before focusing on the field-effect at low temperature. Finally we discuss experiments with other similar devices.

\section{Fabrication of the suspended junctions}

Suspended graphene junctions are fabricated in three main steps. Firstly, a trench is etched into the substrate and a metallic gate is deposited at the bottom of the trench. Secondly, a graphene monolayer is exfoliated, transferred and placed across the trench and above the gate. In the last step, electrodes are patterned to electrically contact the graphene flake on both sides of the trench.

\subsection{Fabrication of the local metallic gate}

\begin{figure}[h]
\begin{center}
\includegraphics[width=14cm]{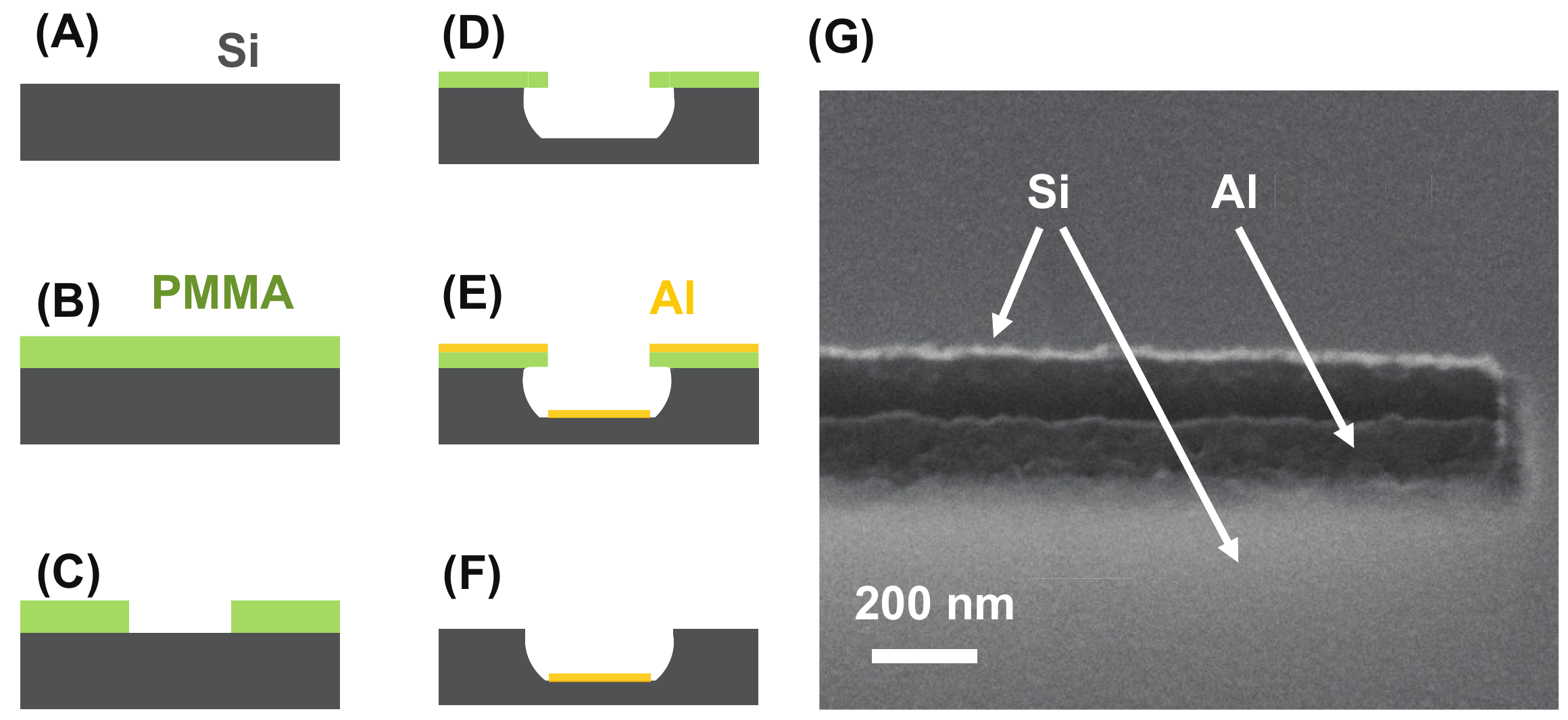}
\vspace{0cm} \caption{(color online) Fabrication of the local gate. (A) Clean bare silicon substrate. (B) Spinning of electronic beam resist (PMMA). (C) Electron beam lithography and development. (D) Dry etching of silicon. (E) Metal evaporation. (F) Lift-off. (G) Scanning electron micrograph (tilted by 45$^\circ$) of a gate after lift-off.}
\label{fig-fab-gate}
\end{center}
\end{figure}

\Fref{fig-fab-gate} describes the successive steps of fabrication of the gate. \Fref{fig-fab-gate}(A): The fabrication starts with a high resistivity silicon substrate (resistivity higher than 10~k$\Omega$.cm). \Fref{fig-fab-gate}(B): The substrate is coated with a layer of polymethylmethacrylate (PMMA) resist. The resist will be used both as an etching mask and as a lift-off sacrificial layer. \Fref{fig-fab-gate}(C): The trench profile is patterned by electron-beam lithography (EBL). \Fref{fig-fab-gate}(D): Isotropic reactive ion etching is performed to etch away 230~nm of silicon. The etching process is carried out using SF$_6$ with a pressure of 50~mTorr, and a plasma created by 50~W of RF power and 300~W of inductive coupling power. \Fref{fig-fab-gate}(E): The etching process creates an undercut, allowing for the deposition of a metallic layer at the bottom of the trench without covering the side walls. A 50~nm layer of aluminum is evaporated to form the gate electrode. \Fref{fig-fab-gate}(F): Lift-off is performed.

\Fref{fig-fab-gate}(G) shows an image  of a finished gate at an observation angle of 45$^\circ$ obtained by scanning electron microscopy (SEM). Trenches and gates with a width as small as 200~nm have been successfully obtained with this process.

\subsection{Transfer of graphene monolayers above the gate}

Graphene monolayers are transferred above the gate using a modified version of the method described in Ref.~\cite{schneider2010}, where cellulose acetate butyrate (CAB) is used as the transfer medium. To allow for an alignment with a precision of a few micrometers, the critical steps of the transfer are performed with the help of a micromanipulator based on a probe station originally designed for electrical measurements. The micromanipulator features a stage with horizontal and vertical translation control, a microscope, and two needle pins with XYZ control.

The graphene transfer step proceeds as follows. First, after the local gate is fabricated, the substrate (from now referred to as the destination substrate) is covered with PMMA. Next, electron beam lithography is performed to open a window of $100~\times~100$~$\mu$m$^2$ centered on the location where graphene is to be transferred. The PMMA acts as a protective layer as the transfer process would otherwise contaminate the destination substrate with graphite residues.

\begin{figure}[h]
\centering
\includegraphics[width=16cm]{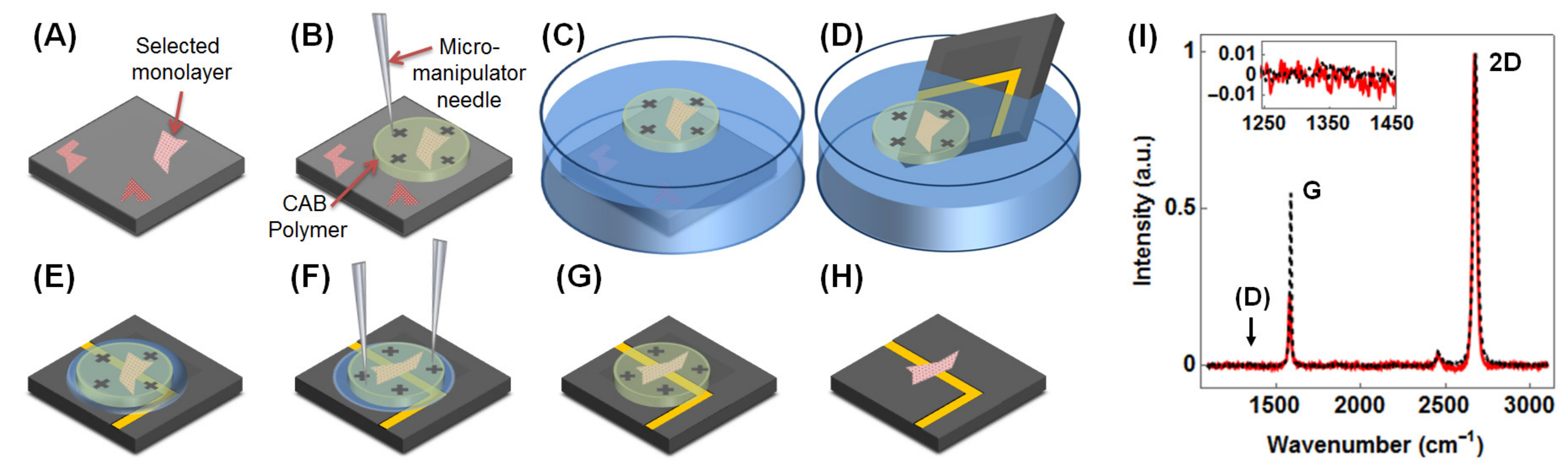}
\vspace{0cm}
\centering
\caption{(color online) Graphene monolayer transfer process. (A) Graphene is mechanically exfoliated on a Si/SiO$_2$ substrate and a monolayer is selected. (B) A drop of CAB polymer covers the selected monolayer. After the polymer is dried markers are patterned using the needles of a micromanipulator. (C) The CAB polymer peels off in deionized water, holding the graphene flake on its bottom side. (D) The CAB polymer is brought on top of the destination substrate. (E) The CAB polymer lies on the destination substrate with a water interface layer. (F) The micromanipulator needles hold the CAB polymer in place whereas the substrate is positioned with the micromanipulator stage. (G) The substrate is baked to evaporate the water film. (H) The CAB polymer and the protecting PMMA resist are dissolved, leaving the transferred monolayer on the destination substrate. (I) Raman spectra of a graphene monolayer before (black dashed line) and after (full red line) transfer. Both spectra were normalized so that the 2D peak maximum intensity is 1 (in arbitrary units). The intensity ratios between the G and 2D peaks are different because of the different substrates \cite{wang2008raman}. The inset zooms on the region where a D peak would appear if the graphene had been damaged by the transfer.}
\label{fig-transfer}
\end{figure}

Graphene layers are then obtained by mechanical exfoliation of kish graphite onto a separate Si/SiO$_2$ substrate (\fref{fig-transfer}(A)), which we call the exfoliation substrate. Using Si/SiO$_2$ as a substrate enables the discrimination of one to few layer flakes by inspection with an optical microscope \cite{blake2007}. The number of layers is then confirmed by Raman spectrometry \cite{ferrari2006}, and a monolayer is selected for the transfer step to follow. A solution of cellulose acetate butyrate (CAB) in ethyl acetate 60~mg/mL, which is a hydrophobic polymer, is  dripped onto the substrate, covering the graphene monolayer of interest. The latter is still distinguishable under the optical microscope after the CAB polymer dries off. The needles of the micromanipulator are used to punch markers around the monolayer (\fref{fig-transfer}(B)). These markers will later  allow for the alignment of the graphene relative to the gate on the destination substrate, onto which graphene is optically indiscernible.

Next, the exfoliation substrate is slowly dipped into deionized water (\fref{fig-transfer}(C)). Since the SiO$_2$ substrate is hydrophilic while the CAB polymer is hydrophobic, water peels off the CAB polymer away from the substrate, whereas the graphene remains attached to the bottom of the polymer \cite{schneider2010}. The polymer is then laid on top of the destination substrate (\fref{fig-transfer}(D)). A film of water with thickness controlled by suction with a pipette separates the polymer from the destination substrate, allowing for the alignment of one relative to the other (\fref{fig-transfer}(E)).

The alignment is achieved using the micromanipulator with needles poking into the polymer and holding it at a fixed position, while the substrate, which is attached to the stage, is displaced (\fref{fig-transfer}(F)). The alignment is performed with the help of the markers previously defined onto the polymer. The precision of this alignment procedure is approximately 5~microns, which is sufficient given the typical graphene monolayer size. The substrate is then baked at 80$^\circ$C for about 10~minutes to evaporate the water film (\fref{fig-transfer}(G)). After baking, the CAB polymer and the underlying protecting PMMA layer are dissolved in acetone (\fref{fig-transfer}(H)). The device is further cleaned in acetone and isopropanol and finally blow-dried with nitrogen gas. In contrast to many fabrication methods of suspended graphene devices \cite{bolotin2008,lau2012}, our process does not require critical point drying.

The transfer area is next imaged by SEM at low electron dose (less than 1~$\mu$C$/$cm$^2$ at 10~keV) to prevent damaging the graphene layer \cite{teweldebrhan2009,murakami2013,childres2010}. The purpose of this observation is to ensure that the graphene overlaps with the trench continuously, and that large enough areas of graphene are present on both sides of the trench, minimizing the contact resistance with the electrodes to be patterned. Furthermore, precise measurements of the position of the graphene flake are performed to prepare the alignment of the next EBL step (patterning of electrodes).

Finally, Raman spectrometry is performed to check the quality of the graphene monolayer after transfer (\fref{fig-transfer}(I)). The absence of D peak in the Raman spectrum ensures that the graphene has not been degraded during the transfer process followed by SEM observation \cite{teweldebrhan2009,murakami2013,childres2010}.

\subsection{Patterning of electrodes}

In the last step of the fabrication process, contact electrodes are patterned using EBL.  The substrate is coated with a PMMA monolayer. After the resist is exposed and developed, 5~nm of titanium and subsequently 50 to 120~nm of aluminum  are evaporated. The final device is obtained after a lift-off step. Again, no critical point drying is required. Electrodes are designed in such a way that at least a few micron squared of graphene are covered on each side of the trench to minimize contact resistance. Ideally, the metal should slightly overlap the trench close to the edge so that the graphene junction area is fully suspended. For the device with $W=7$ $\mu$m, this has not been possible due to lift-off problems. However, we expect that with further improvements in the fabrication process it will be possible to obtain samples where the graphene flake is fully suspended between the contacts. Graphene junctions of length $L$ (cf. figure \ref{fig-sketches}(B)) designed from 300 to 500~nm were successfully fabricated. With this monolayer resist process, lift-off is difficult on devices with shorter junctions. Due to the long aspect ratio of the junctions ($W\geq5\mu$m) a bilayer process is difficult to implement because the resist bridge defining the junction collapses. After the lift-off, no temperature annealing is performed. The device is ready to be measured.

\section{Results and discussion of electrical measurements}

In this section, we present and discuss voltage versus current (V-I) characteristics measured on a typical device. On this sample, the trench is 360 nm wide and the graphene-gate separation is 180 nm. The designed dimensions of the junction considered here are $W=$~7~$\mu$m and $L=$~500~nm. Other devices have been measured; their properties are summarized at the end of this section. As will be described in this section, the devices are not suitable for room temperature use due to the presence of thermal carriers in the substrate. These carriers open a current leakage channel between the gate and electrodes, which masks the regular field effect behaviour. As our device has the specificity to not include a truly insulating layer to separate the gate from the electrodes and the conducting channel, we provide a simple model to quantitatively estimate the leakage currents. This information will help the interested reader to adapt our process to room temperature applications by chosing a more appropriate substrate. The main part of this section focuses on the mobility in the low-temperature regime, where current leakages are suppressed and regular FET behaviour is recovered.

\subsection{V-I characteristics of junctions}

To measure the electric properties of the suspended junctions, the device is glued onto a printed circuit board (PCB) and wire-bonded. The PCB is attached and electrically connected to a sample holder. The sample holder is a copper box shielding the device from electromagnetic noise. Each biasing or measurement port is filtered by a two-stage RC low-pass filter (cutoff frequency ranging from 1~kHz to 2~MHz) in series with a copper powder filter to eliminate higher frequency noise. The sample holder is placed in vacuum in a cryostat enabling measurements from room temperature down to 4~K.

\begin{figure}[h]
\begin{center}
\includegraphics[width=10cm]{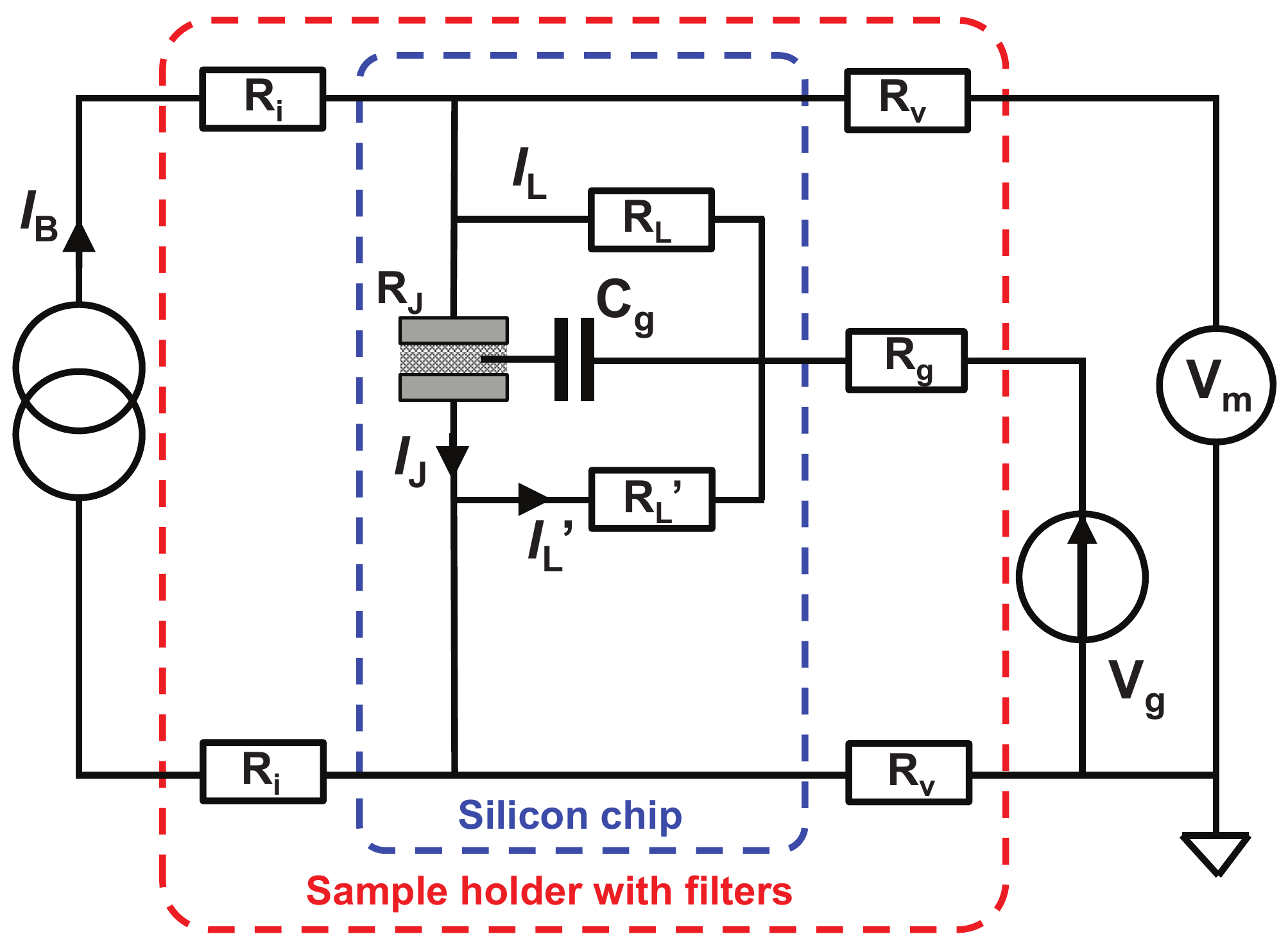}
\vspace{0cm}
\caption{(color online) Electrical circuit modeling the junction in its measurement environment in the presence of gate leakage. The inner dashed (blue) contour encloses the silicon chip space, with parasitic leakage resistances $R_{\rm L}$  and $R'_{\rm L}$. The outer dashed (red) contour represents the sample holder space that includes low-pass filters determined by resistors $R_{\rm i}$ and $R_{\rm v}$ and capacitors (not shown).}
\label{fig-real-circuit}
\end{center}
\end{figure}

Figure \ref{fig-real-circuit} shows the measurement setup. The junction is biased with a current $I_{\rm B}$ delivered by a Yokogawa~7651 source. The voltage drop $V_{\rm m}$ is measured by an Agilent~34401A multimeter. The gate voltage $V_{\rm g}$ is applied between the gate electrode and the ground by another Yokogawa~7651 source used in voltage mode. $R_{\rm i}$, $R_{\rm v}$ and $R_{\rm g}$ are the series resistance added by the filters in the sample holder for the current, voltage and gate voltage leads, respectively. Since only direct current behaviour is of interest here, capacitors of the RC filters are not shown. At the chip level (area enclosed in the blue dashed line on \fref{fig-real-circuit}) the junction is modeled by a resistor $R_{\rm J}$, which is modulated by the gate voltage $V_{\rm g}$ applied through the capacitor $C_{\rm g}$. Unlike the other capacitances, $C_{\rm g}$ is specifically shown for illustration purposes. The resistors $R_{\rm L}$, $R_{\rm L}'$ model parasitic resistances between the gate and source/drain electrodes, as will be discussed in detail below.

\begin{figure}[h]
\begin{center}
\includegraphics[width=16cm]{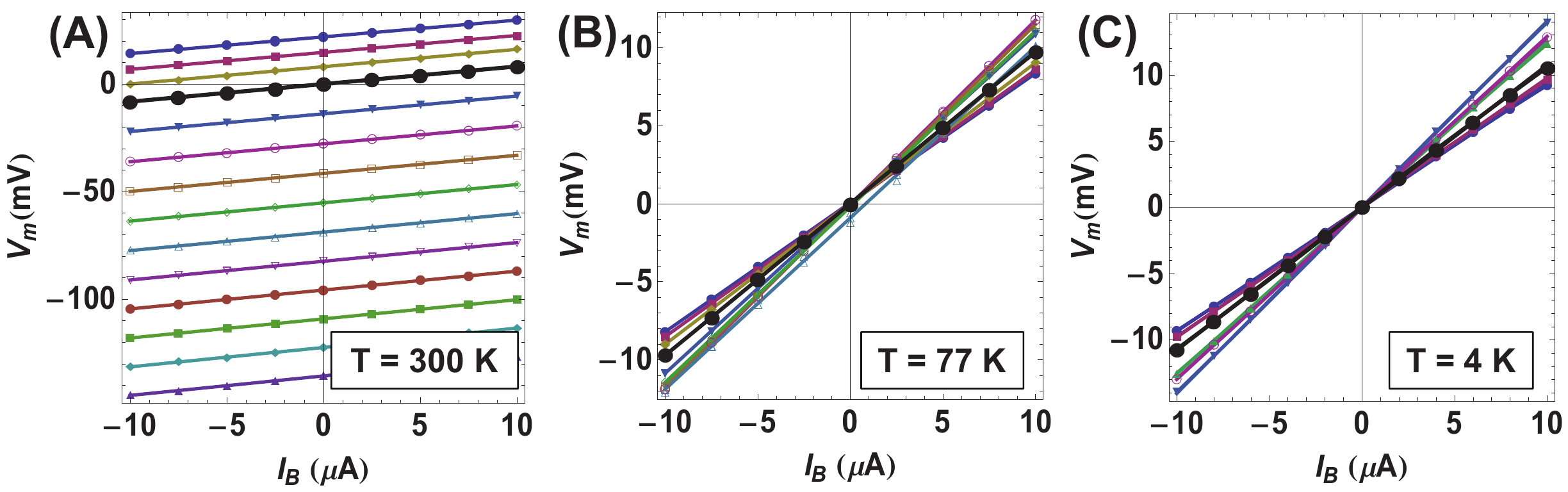}
\vspace{0cm}
\caption{V-I characteristics of the device for various gate voltages $V_{\rm g}$ and 3 different temperatures. On all the panels the symbols are data points, whereas lines are linear fits. The thicker lines with thicker symbols show $V_{\rm m}(I)$ at $V_{\rm g}=0$. (A) $T=300$~K. $V_{\rm g}=$~-10~V (bottom) to 3~V (top) in 1~V steps. (B) $T=77$~K. $V_{\rm g}=$~-5~V to 3~V in 1~V steps. (C) $T=4$~K. $V_{\rm g}=$~-3~V to 2~V in 1~V steps. }
\label{fig-IVs}
\end{center}
\end{figure}

\Fref{fig-IVs} shows three sets of V-Is with varying gate voltages $V_{\rm g}$, corresponding to three different temperatures (300~K, 77~K and 4~K). Symbols are data points, and lines are linear fits.

At room temperature V-Is exhibit a nonconventional behaviour: except for $V_{\rm g}=0$ (larger dots), the V-Is  display a non-zero voltage at zero bias current: $V_{\rm m}(I_{\rm B}=0)\neq 0$. As it will be explained quantitatively in the next subsection, we attribute this offset to gate leakages, \textit{i.e.} a conduction path between the gate and the electrodes. Indeed, unlike common field-effect devices, in our device there is no truly insulating layer (such as silicon oxide or silicon nitride) between gate and electrodes, since both lie on the same substrate. The resistivity $\rho_{\rm Si}$ of our silicon wafers is chosen to be very high ($\rho_{\rm Si}>10$~k$\Omega$.cm) but it is still much smaller than the resistivity of a good insulator (e.g. $\rho_{\rm SiO_2}=10^{14}-10^{16}$~$\Omega$.cm).

At lower temperatures, the V-Is don't show any offset as long as $| V_{\rm g} |$ is kept below a threshold value, typically a few volts. When $| V_{\rm g} |$ is larger than this threshold, the offset kicks in quickly and the V-Is become noisy and non-reproducible. Therefore, we avoid to work in this regime, and the V-Is shown on figures \ref{fig-IVs}.(B) and (C) are confined to a smaller $V_{\rm g}$ range than \fref{fig-IVs}.(A).

\begin{figure}[h]
\begin{center}
\includegraphics[width=14cm]{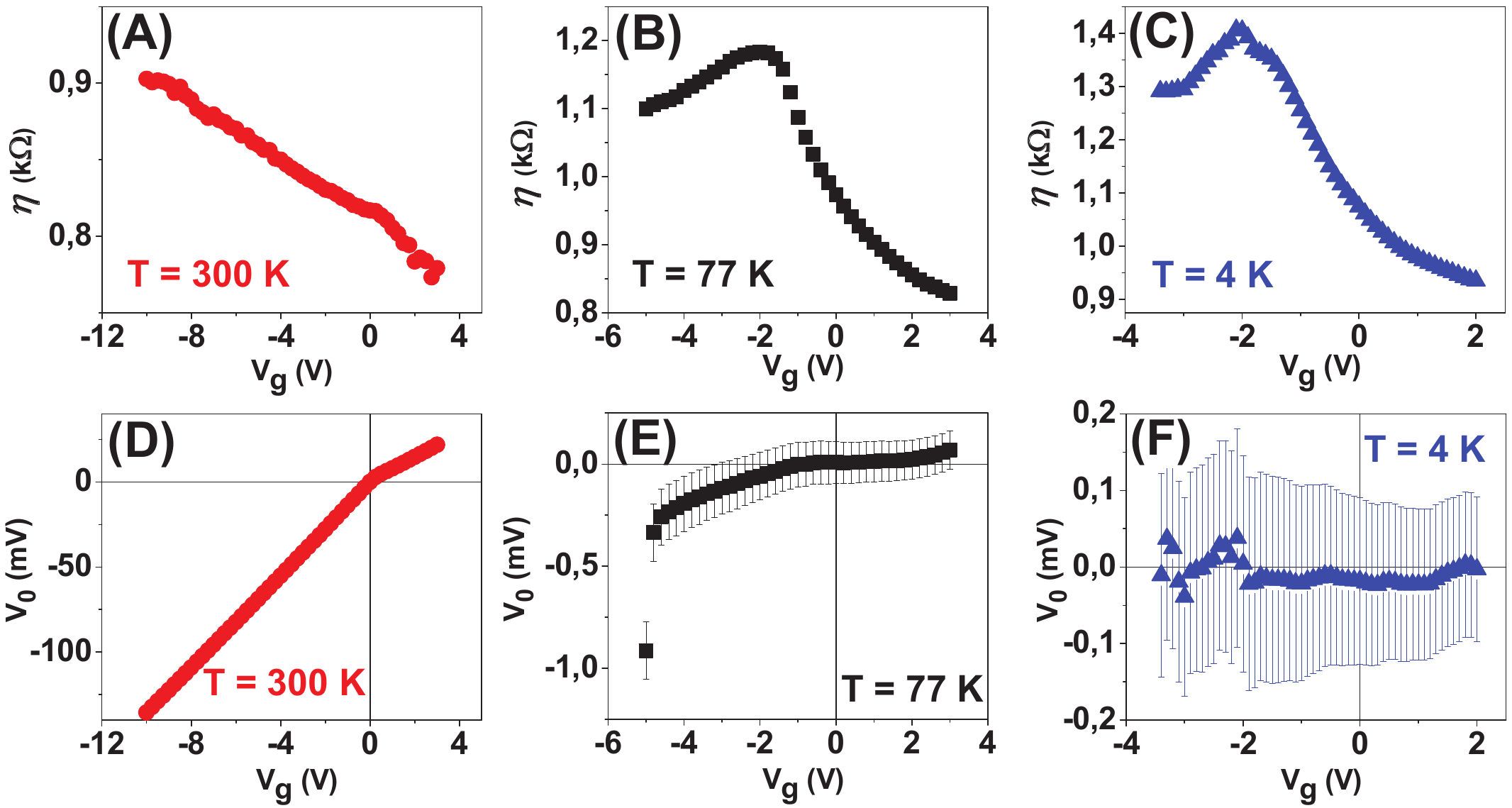}
\vspace{0cm}
\caption{(color online) Fitted slopes $\eta$ (a-c) and vertical offset $V_0$ (d-f) of the V-Is shown on \fref{fig-IVs}. For panels (A), (B), (C) and (D) the error bars are smaller than the symbols.}
\label{fig-eta-V0}
\end{center}
\end{figure}

To allow for a quantitative analysis of the data, we fit each V-I curve to a straight line $V_{\rm m}(I_{\rm B})= \eta I_{\rm B} + V_0$. $\eta(V_{\rm g})$ is the slope in Ohms and $V_0(V_{\rm g})$ is the offset discussed above. \Fref{fig-eta-V0} shows $\eta$ and $V_0$ versus gate voltage $V_{\rm g}$ for the data sets presented on \fref{fig-IVs}. The error bars on $V_0$ have several origins: the standard deviation of the fitting procedure, the accuracy of the voltmeter and the accuracy of the current source.

In an ideal field-effect device the slope $\eta$ of the V-I is equal to the $V_{\rm g}$-dependent junction resistance $R_{\rm J}$. However, because of current leakages through the gate, part of the bias current $I_{\rm B}$ doesn't flow through the junction.  We present in the next subsection a model to take these leakage paths into account.

\subsection{Circuit model}

To quantitatively understand the V-Is presented in the previous subsection we need to account for gate leakages, \textit{i.e.} the finite currents $I_{\rm L}$ and $I'_{\rm L}$ flowing between the gate and the electrodes through the substrate. We model this effect by adding resistors of finite value $R_{\rm L}$ and $R'_{\rm L}$, respectively, between the gate and each electrode (cf. \fref{fig-real-circuit}). In the following we will assume that $R_{\rm L} = R'_{\rm L}$ due to the symmetry of the device. Using Kirchhoff laws, we can calculate the dependence of the fitting parameters $\eta$ and $V_0$ introduced above on the circuit parameters, and so express the junction $R_{\rm J}$ and leakage $R_{\rm L}$ resistances as :

\begin{equation}
R_{\rm J} \simeq \eta,
\label{Rj-exact}
\end{equation}

\noindent and

\begin{equation}
R_{\rm L} \simeq \frac{ (2 R_{\rm v} + \eta) V_{\rm g} - 2 R_{\rm g} V_0}{V_0}.
\label{Rl-exact}
\end{equation}

The expressions above are approximations neglecting terms according to $ R_{\rm v} \ll R_{\rm g}$ and $V_{0}/V_{\rm g} \ll 1$. Before investigating in the next subsection the field-effect properties of the device (variations of $R_{\rm J}$ versus $V_{\rm g}$), we discuss the gate leakage determined based on Eq.\ref{Rl-exact}.

Extracting $R_{\rm L}$ from the fitted $\eta$ and $V_0$ using Eq.\ref{Rl-exact} involves a division by $V_0$. Thus when the uncertainty $\delta V_0$ on $V_0$ is larger than $V_0$ itself, the inferred value for $R_{\rm L}$ is meaningless and should be discarded. On figure \ref{fig-eta-V0} we see that reliable values of $R_{\rm L}$ can be extracted only at room temperature and for a few points at 77~K. At 4~K the leakage resistance cannot be reliably extracted for any value of the presented gate voltage range; however we have also measured V-Is at larger $V_{\rm g}$ (data not shown) so that $R_{\rm L}$ can be estimated.

At room temperature $R_{\rm L}$ is nearly constant  for $V_{\rm g}<0$ with a value of 2~M$\Omega$ which increases to 7~M$\Omega$ as $V_{\rm g}$ is increased from zero to 2~V. We considered a model for the leakage resistance based on two back-to-back Schottky diodes due to the two metal-semiconductor interfaces, and a resistor describing conduction through the silicon substrate. This model leads to a nearly $V_{\rm g}$-independent resistance of 3~M$\Omega$ for $V_{\rm g}<0$, and larger values for $V_{\rm g}>0$ , reaching up to 32~M$\Omega$ for $V_{\rm g}=$~2~V. The agreement is satisfactory given the simplicity of the model and is a strong indication that Schottky barriers and conduction through the substrate are the sources of the measured resistance.

At 77~K and 4~K the leakage resistance is enhanced by at least two orders of magnitude: $R_{\rm L} \approx$ 1 to 2~G$\Omega$ in the $V_{\rm g}$ range where it is computable. In the $V_{\rm g}$ range where $R_{\rm L}$ is not computable, we note that $|V_0|$ tends to be smaller than outside this range. In our parameter range, Eq.\ref{Rl-exact} is dominated by a term proportional to $1/V_0$ so the value of $R_{\rm L}$ computed for $V_{\rm g}$ at the border of the domain of the validity is a lower bound of the actual $R_{\rm L}$ in the inaccessible range. We can then conclude that for $V_{\rm g}\approx 0$, $R_{\rm L} \geq 2$~G$\Omega$ both at 77~K and 4~K. This is consistent with an enhancement of the resistivity of the substrate (intrinsic silicon) and of the Schottky junctions as the temperature is lowered.

In conclusion, gate leakage plays a significant role at room temperature, offsetting the V-Is vertically for arbitrarily small values of the gate voltage $V_{\rm g}$ and disturbing the FET behaviour. However at low temperature ($T \leq 77$ K) gate leakage is strongly suppressed and is negligible over a fairly large range of $V_{\rm g}$. We expect that using as a substrate a higher gap insulator than silicon, for example sapphire, will lead to effective suppression of leakage at room temperature.

\subsection{Field-effect}

Now we analyze the measured $R_{\rm J} (V_{\rm g})$ at low temperature (77~K and 4~K) in terms of field effect. The mobility $\mu$ of the carriers in the graphene junction is given by:

\begin{equation}
\mu = \frac{1}{\rho_{\rm s}~e~n_{\rm tot}},
\label{mu-eq-1}
\end{equation}

\noindent where $\rho_{\rm s}=R_{\rm J} W/L $ is the sheet resistivity of the graphene channel with $R_{\rm J} $ extracted from data using Eq. \ref{Rj-exact}, $-e$ is the electron charge and $n_{\rm tot}$ is the carrier surface density. $n_{\rm tot}=n_{\rm g}+n_{\rm b}$ is the sum of two contributions: $n_{\rm g}$, induced by the gate electric field, and a background $n_{\rm b}$ due to thermally excited carriers as well as charges trapped in the vicinity of the graphene sheet. Using a parallel plate capacitor model,

\begin{equation}
n_{\rm g} = \frac{\epsilon_0 \epsilon_{\rm r} }{t~e} V_{\rm g},
\label{ng-eq}
\end{equation}

\noindent where $\epsilon_0$ is the vacuum permittivity, $\epsilon_{\rm r}=1$ is the relative permittivity of the insulator (here vacuum) and $t=180$~nm is the separation between the gate surface and the graphene sheet. We note that for carrier densities up to $n=2\times10^{11}$ cm$^{-2}$ the attractive electrostatic force between the gate and the graphene bridge does not exceed 11 nN \cite{bolotin2008}, yielding a maximum deflection of 2 nm \cite{fogler2008pseudomagnetic}. Thus we can neglect the dependence of $t$ on the gate voltage in the range explored in the experiment.

\begin{figure}[h]
\begin{center}
\includegraphics[width=15cm]{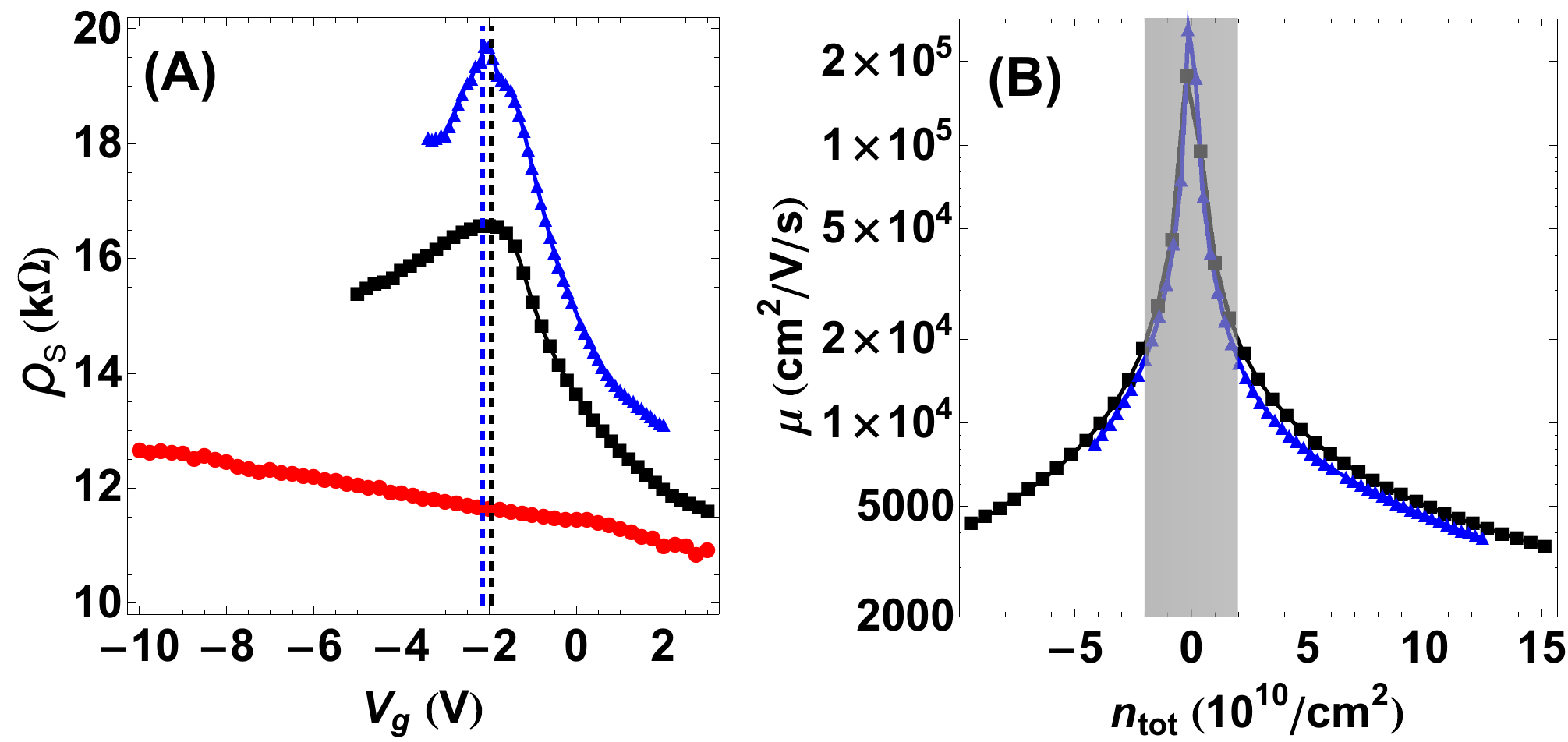}
\vspace{0cm}
\caption{(color online) (A): Sheet resistivity $\rho_{\rm s}$ and (B) mobility $\mu$ extracted from data presented in figure \ref{fig-IVs}, at 300~K (red dots),  77~K (black squares) and 4~K (blue triangles). On panel (A) the vertical lines show the position of the charge neutrality point at 77~K (dashed black lines) and 4 K (dotted blue line). On panel (B) the shaded area masks the domain where Eqs. \ref{mu-eq-1} and \ref{mu-eq-2} do not hold, resulting in an artificial boost of the mobility.}
\label{fig-mobility}
\end{center}
\end{figure}

The other contribution, $n_{\rm b}$, to the carrier surface density is constant and determined by setting the condition $n_{\rm tot}(V_{\rm g}=V_{\rm CNP})=0$. $V_{\rm CNP}$ is the charge neutrality point, \textit{i.e.} the gate voltage for which $R_{\rm J} (V_{\rm g})$ reaches a maximum, determined experimentally (cf. \fref{fig-mobility}(A)). Ideally at $V_{\rm CNP}$ the Fermi level in graphene is set exactly at the point where conduction and valence bands meet, yielding theoretically a zero density of carriers and an infinite resistance. However in practice local fluctuations prevent from nulling exactly the density of carriers \cite{martin2008observation}, and a finite maximum of resistance is observed. In that region $n_{\rm tot}$ becomes independent of $V_{\rm g}$ and Eq. \ref{mu-eq-1} is not valid. Defining $n_{\rm sat}=2.10^{-11}$ cm$^{-2}$ as the total carrier density below which $R_{\rm J}(n_{\rm tot})$ saturates \cite{du2008}, the mobility $\mu$ can be evaluated for $|n|>n_{\rm sat}$ as:

\begin{equation}
\mu = \frac{t}{\epsilon_0  \epsilon_{\rm r} \rho_{\rm s} (V_{\rm g}-V_{\rm CNP})}.
\label{mu-eq-2}
\end{equation}

\Fref{fig-mobility} shows the sheet resistivity $\rho_{\rm s}$ and the mobility $\mu$ for the junction presented earlier. Although the resistivity increases with decreasing temperature, the mobility is rather temperature insensitive. At $n=10^{11}$~cm$^{-2}$, $\mu \approx $~5100~cm$^2$/V/s, which is typical for graphene FETs. In non-suspended \cite{moser2007,ojeda2009} and suspended \cite{bolotin2008,lau2012} graphene junctions however, it has been reported that current annealing can improve the mobility by an order of magnitude. We applied this approach to our junctions. Up to a current density of 1.5~mA/$\mu$m, we indeed observed an increase in mobility, but only marginally (by less than a factor of 2). Above 1.5~mA/$\mu$m, further annealing has no effect or degrades the mobility. Above 3~mA/$\mu$m, all the tested junctions were damaged.  A systematic effect of current annealing though is the reduction of $V_{\rm CNP}$, bringing it closer to zero gate voltage in agreement  with the fact that current annealing cleans the device from charges trapped in the vicinity of the junction \cite{moser2007,cheng2011}.

\begin {table}[h]
\begin{center}
    \begin{tabular}{ | c | c | c | c | c | c | c | }
    \hline
    Device  & $W$  & $L$  & $\rho_{\rm CNP}^{\rm 300K}$ & $\rho_{\rm CNP}^{\rm 77K}$  & $\rho_{\rm CNP}^{\rm 4K}$  & $\mu^{\rm 77K}$ at $n=10^{11}$ cm$^{-2}$ \\
     & ($\mu$m) & (nm) &  (k$\Omega$) &  (k$\Omega$) &  (k$\Omega$) &  (cm$^2$/V/s)  \\ \hline

    W67-8-3 & 7 & 500 & NA & 16.5 & 19.7 & 5100 \\ \hline

    W67-8-2 & 5 & 500 & NA & 10.5 & 10.2 & 11500 \\ \hline

    W67-6-1 & 4 & 350 & 14.5 & 14.0 & NA & 2010  \\ \hline

    \end{tabular}
\caption {Summary of results on measured devices. $\rho_{\rm CNP}$ is the maximum sheet resistivity at the charge neutrality point when it is accessible. $\mu$ is given at 77~K. NA stands for "not available".}
\label{table-results}
\end{center}
\end {table}

Table \ref{table-results} sums up the values of resistivity and mobility measured on three devices. The sheet resistivity at the charge neutrality point $\rho_{\rm CNP}$ at 4~K is given before current annealing. The latter was found to slightly change $\rho_{\rm CNP}$ (up to $+17\%$ and down to $-7\%$), without sizeable effect on the mobility. The mobilities are given at 77~K, but data at 4~K, when available, show that the mobility drops by $\approx 10 \%$ compared to the 77~K value.

The graphene junctions presented in this work have a sheet resistance and a mobility similar to those of state-of-the-art graphene FETs, although their mobility is comparable only to unsuspended devices. However this is sufficient to fabricate graphene-based Josephson junctions \cite{ojeda2009,du2008b}.

Before concluding, we note that related processes to transfer and suspend graphene on top of existing structures were recently published \cite{castellanos2014, weber2014}: in those works the transfer medium is either a PDMS or PMMA film and the electrical contact to the rest of the circuit is performed by stamping this film directly onto metallic electrodes. While these methods may be successful to build graphene FETs, no DC electrical characterization is presented although very promising results were obtained at RF-frequencies \cite{weber2014, singh2014}. Patterning contacts by metal evaporation on top of a graphene layer, as done in our process, is a well established and reliable method to obtain low resistance ohmic contacts required for DC applications, and it allows for a precise definition of the geometry of the contacts.

\section*{Conclusions}

We presented a fabrication method for field-effect transistors made of a graphene monolayer sheet suspended above a local metallic gate and lying on an insulating substrate. In these first experiments, the sheet resistivities and mobilities of the devices at low temperature are comparable to those of unsuspended graphene devices.
The devices described in this work constitute a viable route towards graphene-based superconducting quantum devices. First, the measured mobilities are expected to be sufficient to build graphene Josephson junctions with critical current tunable by the gate voltage. Second, in contrast with graphene devices gated by a doped substrate, using an insulating substrate provides an environment compatible with superconducting qubits and circuit-QED experiments. Indeed, energy loss through the radiofrequency excitation of substrate carriers is suppressed, increasing the coherence time of quantum devices. Finally, local gating enables the independent control of individual graphene junctions on the same chip, which is decisive for scalability.
The fabrication process we describe in this paper is also of interest for investigations of the nano-electromechanical properties of graphene. We finally note that the transfer and suspension method could be applied to fabricate devices incorporating 2D materials other than graphene, including boron nitride or molybdenum disulfide.

\section*{Acknowledgements}

We thank Thomas Szkopek and Elisabeth Ledwosinska for useful discussions, and Martin Otto for help with the experiments. We thank the Nanofab team at IQC, in particular Nathan Nelson-Fitzpatrick, for their advice all along the development of the fabrication process. We are grateful to Liyan Zhao and the Waterloo Advanced Technology Lab for their help with the Raman spectrometer. We thank Pol Forn-Diaz for valuable inputs to the manuscript. This work was supported by the Natural Sciences and Engineering Research Council of Canada, the Canada Foundation for Innovation, the Ontario Ministry of Research and Innovation, and CMC Microsystems. During this work, Adrian Lupascu was supported by an Alfred Sloan Foundation Fellowship.

\section*{References}

\bibliographystyle{iopart-num}
\bibliography{ong-graphene-bibliography}

\end{document}